\documentclass[superscriptaddress,aps,preprintnumbers,amsmath,showpacs,amssymb,prd,nofootinbib,preprint]{revtex4-1}
\pdfoutput=1
\usepackage{booktabs} 
\usepackage{array} 
\usepackage[table,xcdraw]{xcolor} 
\usepackage{amsmath,amssymb}
\usepackage{graphicx}
\usepackage{fancyhdr}
\usepackage{bm, color}
\usepackage{slashed,amsthm,amsfonts,empheq}
\usepackage[caption=false]{subfig}
\usepackage{hyperref}
\usepackage{booktabs}
\usepackage{multirow}
\usepackage[left=2cm,right=2cm,top=2cm,bottom=2cm,a4paper]{geometry}

\newcommand{\Slash}[1]{{\ooalign{\hfil#1\hfil\crcr\raise.167ex\hbox{/}}}}

\newcommand{\beq}{\begin{equation}}  \newcommand{\eeq}{\end{equation}}
\newcommand{\bef}{\begin{figure}}  \newcommand{\eef}{\end{figure}}
\newcommand{\bec}{\begin{center}}  \newcommand{\eec}{\end{center}}
  
\newcommand{\laq}[1]{\label{eq:#1}}  

\newcommand{\Eq}[1]{Eq.(\ref{eq:#1})}

\newcommand{\eq}[1]{(\ref{eq:#1})}

\newcommand{\vev}[1]{\left\langle {#1} \right\rangle}

\def\({\left(}
\def\){\right)}

\newcommand{\EV}{{\rm eV}}

\newcommand{\GEV}{{\rm GeV}}

\def\a{\alpha}

\def\d{\delta}

\def\f{\phi}
\def\g{\gamma}

\def\D{\Delta}

\def\tl{\tilde}
\def\*{\dagger}

\begin{document}
%
%
%
%
\preprint{TU-1268}

\title{
Relativistic Axion with Nonrelativistic Momenta: \\
A Robust Bound on Minimal ALP Dark Matter}


\author{Yuma Narita}
\affiliation{Department of Physics, Tokyo Metropolitan University, Minami-Osawa, Hachioji-shi, Tokyo 192-0397, Japan}
\affiliation{Department of Physics, Tohoku University, Sendai, Miyagi 980-8578, Japan}

 \author{Wen Yin}
\affiliation{Department of Physics, Tokyo Metropolitan University, Minami-Osawa, Hachioji-shi, Tokyo 192-0397, Japan}

\begin{abstract}
The axion‐like particle (ALP), a pseudo Nambu-Goldstone boson that couples to two photons, has been studied extensively in recent years as a dark matter candidate.  
For initial field configurations in a minimal ALP model explaining the observed dark matter abundance, we need the potential height to exceed the ALP energy density at redshift $z\approx 5.5\times10^{6}$ leading to:
$$
    f_{\phi}\gtrsim4\times10^{13}\,\mathrm{GeV}\,\biggl(\frac{10^{-18}\,\mathrm{eV}}{m_{\phi}}\biggr),
$$
where $m_{\phi}$ and $f_{\phi}$ denote the ALP mass and decay constant, respectively.  
This bound is known for the ALP dark matter dominated by the homogeneous zero‐momentum mode, under the requirement that coherent oscillations begin early enough to satisfy the late‐forming dark matter constraint. One loop hole to evade this limit may be to introduce a large amount of the non-relativistic modes of the ALP with non-vanishing momenta. 
Here we show that the same limit remains valid even if nonzero‐momentum modes dominate.  
Interestingly, when {\it nonrelativistic} gradient and kinetic modes prevail, the ALP behaves {\it relativistic} radiation rather than matter, if it violates the limit. Moreover, if the typical momentum is sufficiently small, Baumkuchen-like domain walls form, which play an important role in understanding the transition.

\noindent
\end{abstract}
\maketitle


\maketitle

\section{Introduction} 

The axion or axion-like particle (ALP) is a pseudo Nambu-Goldstone boson whose mass arises from the explicit breaking of its shift symmetry (see Refs.~\cite{Jaeckel:2010ni,Ringwald:2012hr,Arias:2012az,Graham:2015ouw,Marsh:2015xka,Irastorza:2018dyq,DiLuzio:2020wdo} for reviews). A well-known example is the QCD axion, which provides a solution to the strong CP problem via the Peccei-Quinn mechanism~\cite{Peccei:1977hh,Peccei:1977ur,Weinberg:1977ma,Wilczek:1977pj}.

Dark matter provides direct evidence for physics beyond the Standard Model, yet its nature remains unknown. Among the proposed candidates, the axion is particularly well motivated. In the misalignment mechanism~\cite{Preskill:1982cy,Abbott:1982af,Dine:1982ah}, the axion begins coherent oscillations when its mass becomes comparable to the Hubble parameter, and the resulting energy density can account for dark matter. It has also been shown that if inflation lasts sufficiently long and the Hubble parameter during inflation is lower than the QCD scale, the predicted axion abundance can be significantly altered~\cite{Graham:2018jyp,Guth:2018hsa,Ho:2019ayl,Alonso-Alvarez:2019ixv}.
 Additionally, the axion/ALP can be initially placed near the hilltop of its potential through interactions with other fields, thereby delaying the onset of oscillations and enhancing its abundance~\cite{Daido:2017wwb,Co:2018mho,Takahashi:2019pqf,Nakagawa:2020eeg,Narita:2023naj}. Light ALPs can also be produced via stimulated emission, such as from inflaton, Higgs decay or even thermal scattering~\cite{Moroi:2020has,Moroi:2020bkq,Nakayama:2021avl,Choi:2023jxw,Yin:2023jjj,Sakurai:2024apm}, or through wide resonance effects~\cite{Co:2017mop,Harigaya:2019qnl}.

More recently, it has been shown that a first-order phase transition relevant to the mass generation can significantly enhance ALP dark matter production, leading to novel dynamics associated with bubble expansion and axion wave generation~\cite{Nakagawa:2022wwm,Lee:2024oaz}. In addition to these scenarios, various other mechanisms for ALP and QCD axion dark matter production have been extensively studied for the dark matter valid in wide parameter region. Experimental efforts are also ongoing for the large discovery space; see, e.g., Ref.~\cite{AxionLimits}.

In this paper we focus on a minimal ALP model.  In this model the ALP potential consists of a single cosine term:
\beq\laq{pote}
V = m_\f^2 f_\f^2\bigl(1 - \cos\bigl(\f/f_\f\bigr)\bigr).
\eeq
This potential is bounded above and below, which gives an  constraint on ALP dark matter from the onset of coherent oscillations, as shown in, e.g., Refs.\,\cite{Marsh:2019bjr,Dror:2020zru,Nakagawa:2022wwm} by taking account of constraints for the small scale structure from Lyman-$\alpha$ and Milky Way subhalo count.

Observations of small‐scale structure place a lower bound on the redshift for the ALP being the matter~\cite{Sarkar:2014bca,Corasaniti:2016epp,Das:2020nwc}.  In particular, the abundance of Milky Way satellite galaxies sets the lower bound \cite{Das:2020nwc,Fujita:2023axo}
\beq\laq{mat}
z_{\rm matter} \gtrsim 5.5 \times 10^6.
\eeq
Requiring the potential hight is dominant
\beq 
\laq{limit1}
(m_\phi f_\phi)^2 > \rho_\phi,
\eeq
with the total energy density of the ALP, $\rho_\phi = \rho_{\rm DM}\simeq (1+z)^3(1.8\times10^{-12}\,\GEV)^4$ at $z=5.5\times10^6$, yields~\cite{Marsh:2019bjr, Nakagawa:2022wwm}
\beq\laq{limit}
f_\phi \gtrsim 4\times10^{13}\,\GEV \biggl(\frac{10^{-18}\,\EV}{m_\phi}\biggr),
\eeq
by assuming the ALP composed mostly by the homogeneous mode behaving like pressureless matter. 

In this work, we demonstrate that the same limit applies for even  more generic initial field configurations, where nonzero-momentum modes dominate the energy density.  
It is clear that relativistic ALP particles cannot violate limit \eq{limit}, because it is obviously not the matter. 
The only possible loophole to violate the limit \eq{limit} may be the non-homogenous but non-relativistic particles. 
However, interestingly, we find through lattice simulations that when the ALP is dominated by nonrelativistic gradient modes, it nevertheless behaves like relativistic radiation if the condition~\eq{limit1} is violated.  
Furthermore, if the typical momentum is sufficiently small, the field can form domain walls once the gradient energy redshifts and becomes comparable to the potential energy.  
This implies that the constraint~\Eq{limit} is robust.  Assuming the $O(1)$ anomaly coefficient to the photon, we derive a stringent bound to the minimal ALP dark matter model. 

We comment on that a generic limit on the dark photon with mass given by a dark Higgs field was pointed out. This is because if the coupling is too large, the back reaction of the abundant dark photon  brings the dark Higgs into symmetric phase~\cite{Kitajima:2024jfl} (see also a parametric similar limit from topological defect formation~\cite{Cyncynates:2023zwj, Cyncynates:2024yxm}). This paper will show a similar result to ALP from different dynamics: if the coupling $1/f_\f$ is too large, the existence of too much ALP fluctuation makes the potential effectively flat.

\section{Relativistic axion with non-relativistic momenta}

Here we perform a lattice simulation by using {\tt Cosmolattice}~\cite{Figueroa:2020rrl,Figueroa:2021yhd} to study the impact on the non-homogeneous mode given that the potential \Eq{pote} is a cosine term, which is non-linear in $\f$. {We will show when \Eq{limit1} is satisfied, even if the momentum is smaller than $m_\f$ the ALP behaves as radiation. }

Let us consider 
\beq
\f_i = \bar{\f}_i + \delta \f_i,
\eeq
where $\bar{\f}_i$ is the homogeneous mode, that we take $\bar \f_i=0, \dot{\bar{\f_i}}=0$, 
since the ALP limit on the dark matter with the dominated homogeneous mode is known. 
$\dot X \equiv d X/d t $ with $t$ being the cosmic time.  
 $\delta \phi$ represents the fluctuation.
Throughout this paper, we use the subscript $i$ to denote the initial condition for the simulation. 

The fluctuation $\d \f_i $ satisfies
\beq
\langle \delta \f_i \delta \f_i \rangle =  \int^{K_{\rm UV}}_{K_{\rm IR}} \frac{d k}{k} \, {\cal P}_{\delta \f}(k).
\eeq
Here 
$K_{\rm UV, IR}$ is the momentum cutoff for the spectrum we consider. 
{Namely, we assume the reduced power spectrum vanish out of the range $[K_{\rm IR}-K_{\rm UV}]$. }
We take 
 $K_{\rm IR}\gtrsim H_i$, with $H$ being the Hubble parameter, 
 because we consider the fluctuation is adiabatic for conservativeness, and we choose the uniform density slicing for the super-horizon mode. 
{This is the case for instance the ALP is produced from inflaton decay. In this case, we will not have the constraint} from large scale isocurvature modes. 

{
To ensure a vacuum-like state with minimal energy, the field fluctuation and its time derivative are related by the on-shell condition,
\begin{equation}
   \delta\dot{\phi}_{i,\vec{k}}
      = \left( \pm i \omega_k - aH \right)\, \delta\phi_{i,\vec{k}} ,
\end{equation}
where $\delta\phi_{i,\vec{k}}$ and $\delta\dot{\phi}_{i,\vec{k}}$ denote the Fourier modes of 
$\delta\phi$ and $\delta\dot{\phi}$, respectively.  
The frequency is given by
\begin{equation}\laq{inw}
\omega_k = \sqrt{\vec{k}^{\,2} + m_\phi^{\,2}},
\end{equation}
following the standard vacuum prescription used in lattice simulations such as {\tt CosmoLattice}.  
Here $m_\phi$ is the curvature of the potential evaluated at the homogeneous background $\bar{\phi}$; in our setup we take $\bar\phi = 0$.  
This initialization corresponds to the usual dispersion relation for ALP ``particles''.  
Simulations with a different choice for the initial condition are shown in Appendix~\ref{app:3}.
}

For the reduced power spectrum, we take the form 
\beq
{\cal P}_{\delta \f}(k) = \tilde{\phi}^2 \(\frac{k}{m_\f}\)^q,
\eeq
with $\tl \f$ being a normalization.  
It is white noise with $q=3$ and scale invariant for $q= 0$. 
For $q\neq 0$, this gives 
\beq
\langle \delta \f_i \delta \f_i \rangle= \frac{ \tilde{\phi}^2 }{q}  \((\frac{K_{\rm UV}}{m_\f})^q-(\frac{K_{\rm IR}}{m_\f})^q\),
\eeq
while the $q\to0$ limit gives the case for $q=0$.

To study the late time cosmology, we solve the equation of motion,
\beq\laq{eom}
\ddot{\f}-\frac{1}{a^2}\Delta\phi + 3H \dot{\f}  = -\partial_{\f}V,
\eeq 
by assuming a radiation dominated Universe for our purpose. $a$ is the scale factor, which is taken to be unity, $a_i=1$, at the initial simulation time. {$\D$ is the Laplacian for the comoving coordinate.}

\begin{figure*}[!t]
    \begin{center}
        \includegraphics[width=1\textwidth]{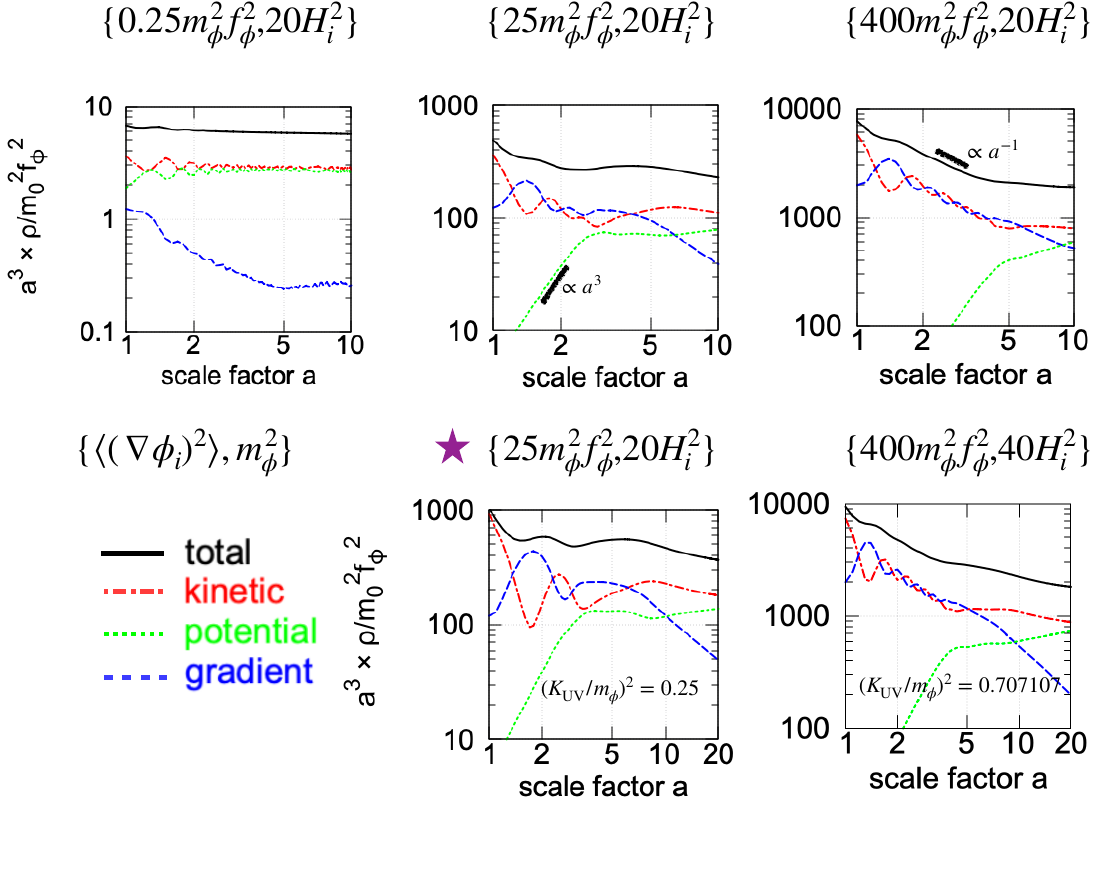}
        \vspace{-15mm}
    \end{center}
    \caption{
    Time evolution of the energy density normalized by $a^3$. 
    The red dot-dashed, blue dashed, green dotted, and black solid lines represent the kinetic, gradient, potential, and total energy densities, respectively. 
    In all panels we set $H_i = 0.5\,m_\star$, $q = 2$, $K_{\rm IR} = H_i$, and a comoving box size $L \simeq 63/m_\star$ with $a_i = 1$. 
    The top-left, top-middle, top-right, bottom-middle, and bottom-right panels correspond to 
    $(\tilde{\phi}, K_{\rm UV}) = (f_\phi, m_\phi)$, $(10 f_\phi, m_\phi)$, $(40 f_\phi, m_\phi)$, $(40 f_\phi, 0.5 m_\phi)$, and $(40 f_\phi, \approx 0.84 m_\phi)$, respectively. 
    In all panels we take $m_\phi^2 = 5 m_\star^2$, except for the bottom-right one, where $m_\phi^2 = 10 m_\star^2$. {The bold black segments are added for visual reference to check the scaling behavior. 
The purple star in the bottom-middle panel indicates the point corresponding to the plots shown in Figs.~\ref{fig:2} and \ref{fig:5}.} }
    \label{fig:1}
\end{figure*}

Fig.~\ref{fig:1} shows the time evolution of the energy density normalized by $a^3$. 
The red dot-dashed, blue dashed, green dotted, and black solid lines represent the kinetic, gradient, potential, and total energy densities, respectively. 
In all panels we set $H_i = 0.5\,m_\star$, $q = 2$, and $K_{\rm IR} = H_i$. 
The five panels correspond to different choices of $(\tilde{\phi}, K_{\rm UV})$, namely 
$(f_\phi, m_\phi)$, $(10 f_\phi, m_\phi)$, $(40 f_\phi, m_\phi)$, $(40 f_\phi, 0.5 m_\phi)$, 
and $(40 f_\phi, \simeq 0.84 m_\phi)$, from the top-left to the bottom-right. 
In all panels we take $m_\phi^2 = 5 m_\star^2$, except for the bottom-right one, where $m_\phi^2 = 10 m_\star^2$. 
Here $m_\star$ denotes the machine unit.\footnote{For other parameter sets, our conclusions remain qualitatively unchanged. 
For $-2 < q < 0$, the results differ quantitatively but not qualitatively.}

{The top-left panel shows the case in which \Eq{limit1} is satisfied, 
whereas in the other four panels \Eq{limit1} is violated. 
This can be understood from the initial gradient energy density,
\beq
\frac{1}{2}\vev{(\nabla \phi_i)^2} = \frac{1}{2}\tilde{\phi}^2
\frac{K_{\rm UV}^{2+q}/m_\phi^q - K_{\rm IR}^{2+q}/m_\phi^q}{2+q},
\eeq
which reduces to
\beq
\frac{1}{2}\vev{(\nabla \phi_i)^2} \sim  \frac{1}{8} \frac{\tilde{\phi}^2 K_{\rm UV}^4}{m_\phi^2}.
\eeq
The kinetic energy is estimated 
\beq
\frac{1}{2}\vev{(\dot{\f}_i)^2}\sim \frac{m_\f^2}{2} \frac{ \tilde{\phi}^2 }{q}  \((\frac{K_{\rm UV}}{m_\f})^q-(\frac{K_{\rm IR}}{m_\f})^q\)\sim \frac{1}{4} \tilde{\phi}^2  K^2_{\rm UV},
\eeq
where we used that $K_{\rm UV}\lesssim m_\f,$ and approximated $\omega \approx m_\f.$ This is comparable to or larger than the gradient energy when $K_{\rm UV}\lesssim m_\f.$}

For the parameter choices in Fig.~\ref{fig:1}, 
the initial values of $\vev{(\nabla \phi_i)^2}$ are 
$\frac{1}{4} m_\phi^2 f_\phi^2$, $25 m_\phi^2 f_\phi^2$, 
$400 m_\phi^2 f_\phi^2$, $25 m_\phi^2 f_\phi^2$, and $400 m_\phi^2 f_\phi^2$, 
from the top-left to the bottom-right panels, respectively.  These values are also indicated at the top of each panel.\footnote{
The simulations in the top panels employ $512^3$ lattices, while those in the bottom panels use $1024^3$ lattices.}

We clearly see that the total energy in the top-left panel behaves as matter, 
whereas in the other panels it initially behaves like almost radiation.
The potential energy in the top-left panel scales as matter, 
while in the other cases it first behaves as a cosmological constant, 
$\vev{V} a^3 \propto a^3$, and later transitions to matter-like behavior. 

This difference can be intuitively understood as follows. 
For the relevant modes $k_{\rm IR} < k < k_{\rm UV} \le m_\phi$ with large amplitudes, 
the right-hand side of \Eq{eom}, bounded by $m_\phi^2 f_\phi$, 
can be neglected. In this regime, the gradient energy becomes comparable to the kinetic energy, 
and the ALP field behaves as radiation, i.e., the massless limit. 
The averaged potential energy is then $\vev{V} = m_\phi^2 f_\phi^2$, 
which effectively behaves as a cosmological constant. 

{As the gradient and kinetic energies redshift and become comparable to the potential energy, 
a transition in the equation of state occurs in the top-middle to bottom-right panels. 
Analytically, this transition happens at {$a \sim 2.5$, $ 4.9$, $3.3$, and $4.4$, for the top-middle, top-right, bottom-middle, and bottom-right,} respectively.\footnote{{For discussing the limit in \Eq{mat}, this period can be regarded as corresponding to $z_{\rm matter}$. 
At this stage, $m_\phi$ is about an order of magnitude larger than $H$. 
Although increasing this hierarchy is computationally challenging in lattice simulations, 
we expect that our results would not change significantly for a larger hierarchy, 
as can be inferred from the comparison between the top-right and bottom-right panels in Fig.~\ref{fig:1}. }}
The figure indeed shows that the behavior changes at these epochs, 
particularly in the potential energy. 
By comparing the top and bottom panels, 
one can see that the transition epoch depends only weakly on 
$K_{\rm UV}$, $\tilde{\phi}$, or $m_\phi$, 
as long as the initial {total energy density} is fixed. }

\begin{figure}[!t]
    \begin{center}
        \includegraphics[width=145mm]{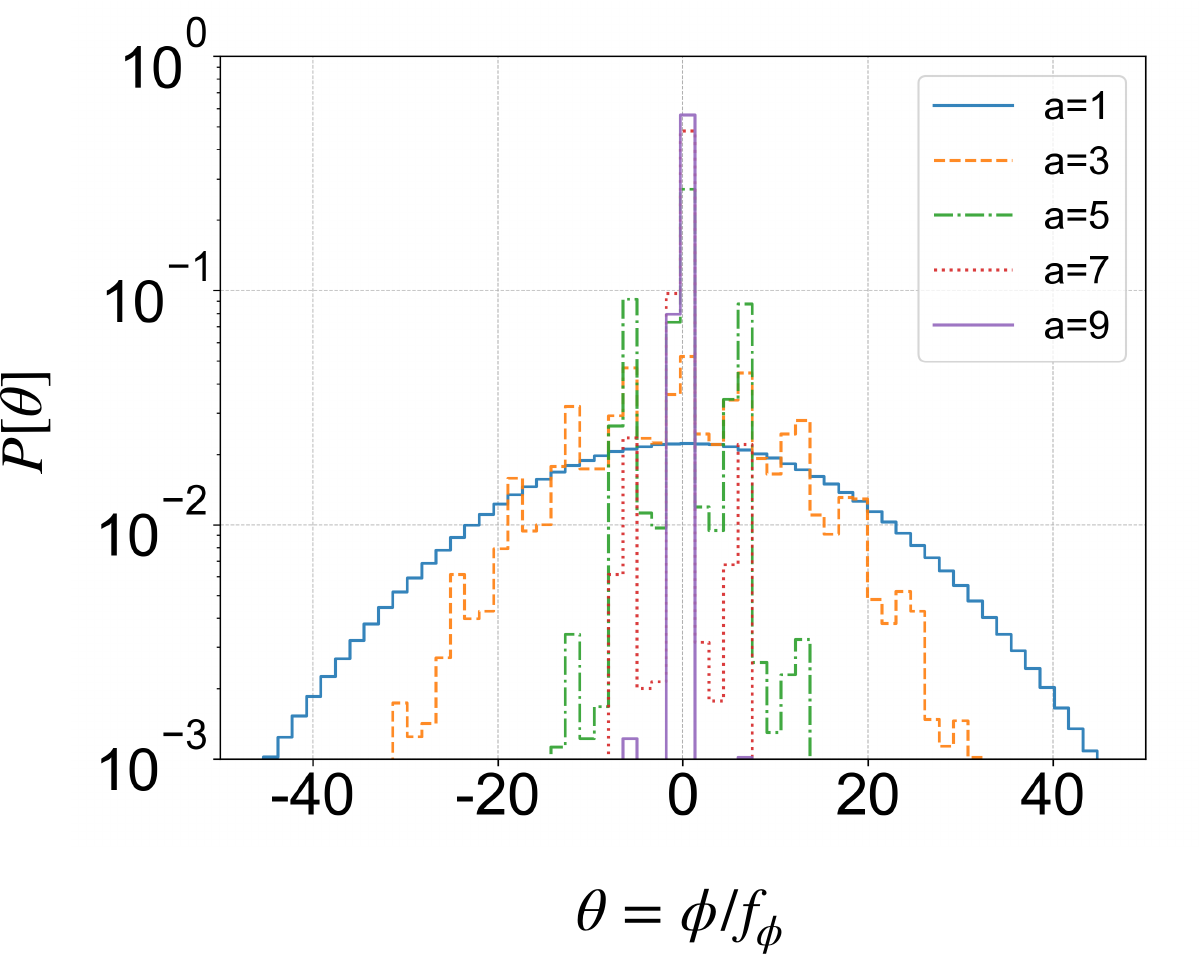}
        \includegraphics[width=145mm]{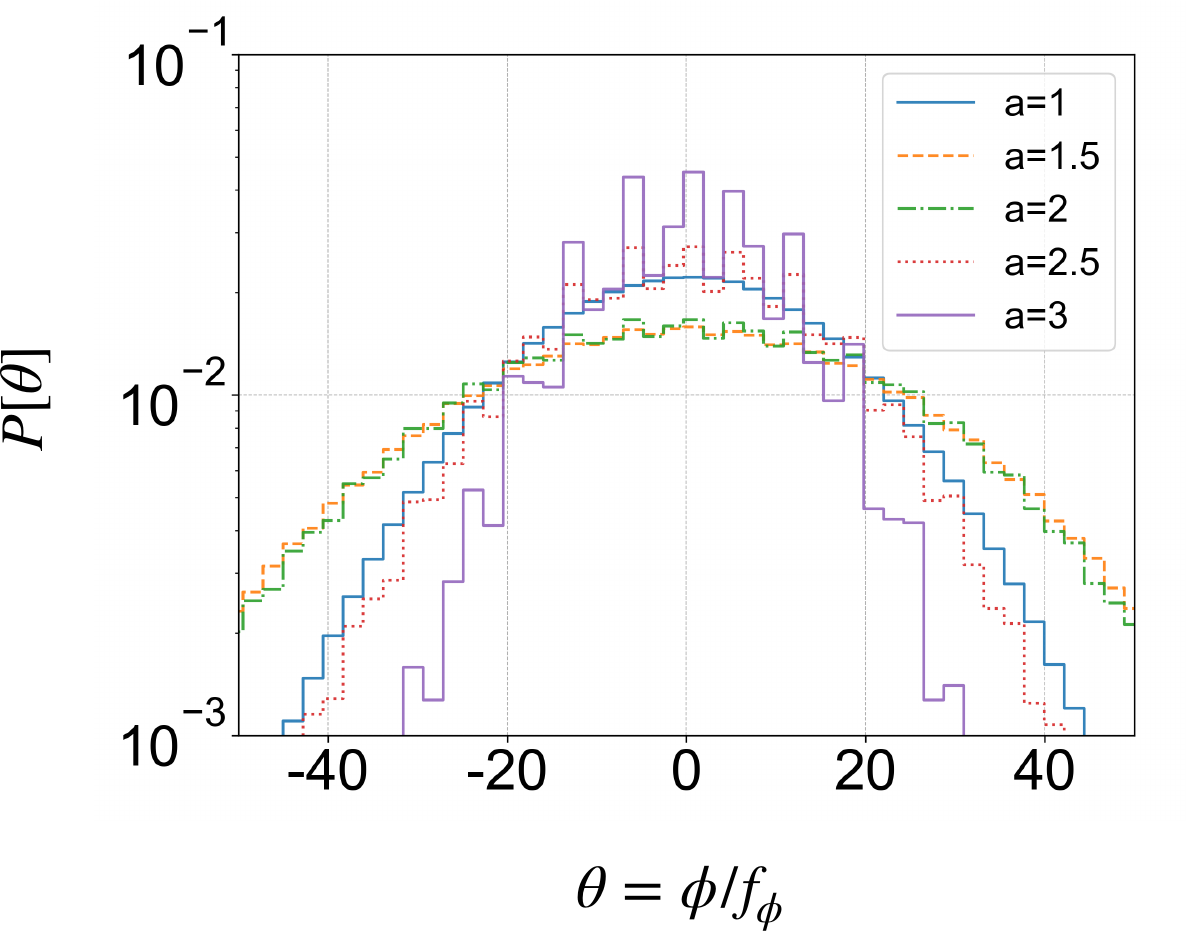}
        \vspace{-5mm}
    \end{center}
    \caption{
    Distribution of $\phi/f_\phi$ for the bottom-middle panel of Fig.~\ref{fig:1}. 
    The lower panel focuses on the short time interval until $a= 3$ (c.f. the transition at $a\sim 2.5$).}
    \label{fig:2}
\end{figure}

\begin{figure}[!t]
    \begin{center}
        \includegraphics[width=145mm]{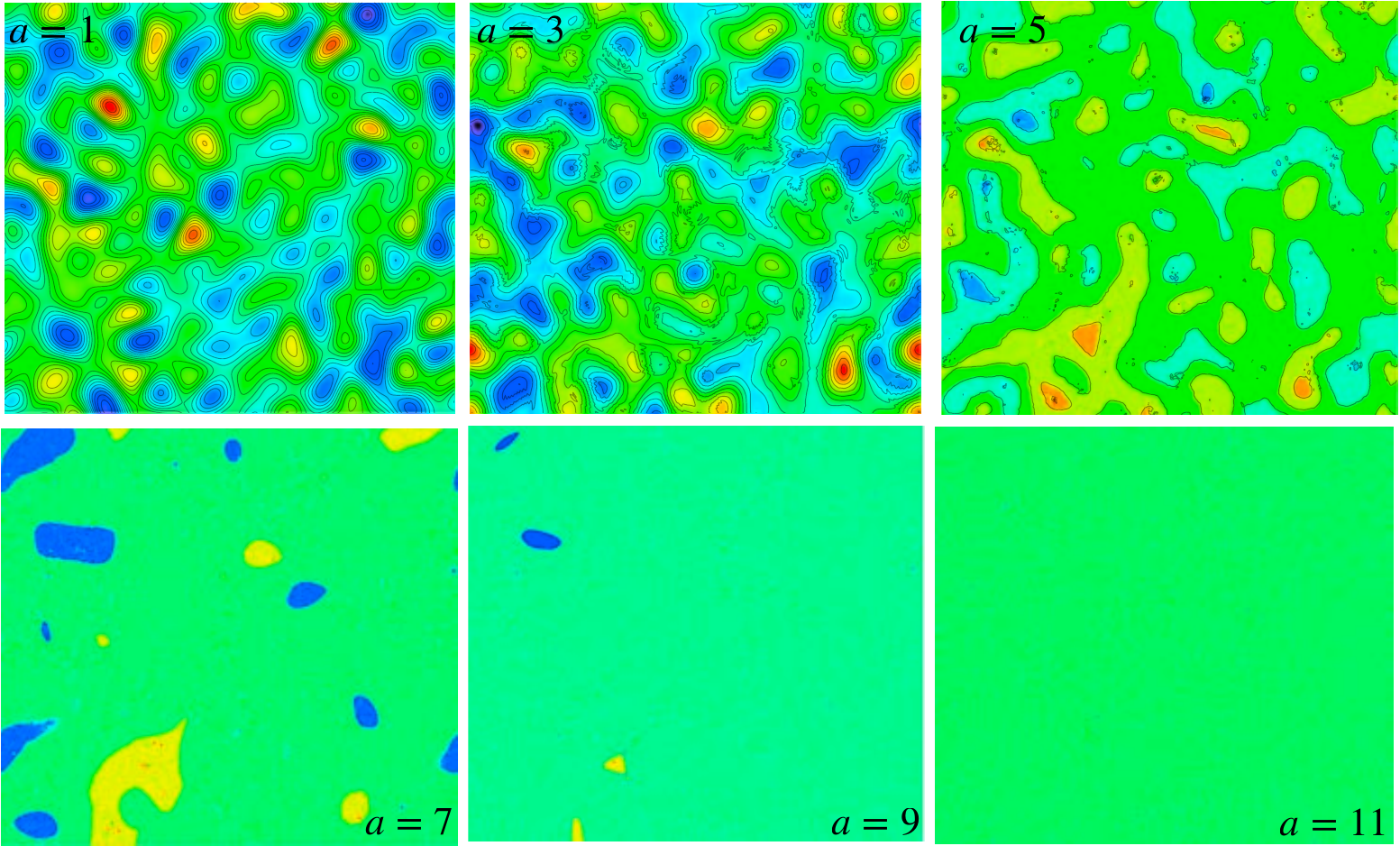}
        \vspace{-5mm}
    \end{center}
    \caption{
    Snapshots of $\phi/f_\phi$ on an $x$-$y$ slice 
    (we choose the 513th slice along the $z$ direction) 
    at $a = 1, 3, 5, 7, 9,$ and $11$. {The black contours in the top panels mark the locations of the domain walls.}}
    \label{fig:5}
\end{figure}

\paragraph*{Relation of transition to formation of Baumkuchen domain wall}
Note that when this transition occurs with nonrelativistic momenta, $H<k_{\rm typ} < m_\phi$, 
the typical field amplitude $|\delta\phi_{\rm amp}|$ satisfying 
$\vev{(\nabla \phi)^2} \sim k_{\rm typ}^2 \delta\phi_{\rm amp}^2 \sim m_\phi^2 f_\phi^2$ 
is larger than $f_\phi$. 
Thus, the field configurations in coordinate space cross multiple potential barriers, and 
domain walls are formed around or slightly before the transition epoch. 
Due to the population bias among different vacua, 
we expect the domain walls to collapse within a relatively short lifetime.

The histogram of $\phi$ corresponding to the bottom-middle panel of Fig.~\ref{fig:1} 
is shown in the top panel of Fig.~\ref{fig:2} (see also the other simulations in Appendices \ref{app2} and \ref{app:3}). 
It clearly indicates that the field values become localized around multiple vacua, 
$\phi = 2 n \pi f_\phi$ with integer $n$, soon after $a \gtrsim 2.5$, 
confirming that domain walls are indeed formed by field configurations connecting these vacua. 
Subsequently, the domain wall network collapses into a single domain around $\phi / f_\phi \simeq 0$, 
as seen for $a \to 9$. 

Together with the bottom panel of Fig.~\ref{fig:2}, which focuses on the short period {$a \lesssim 3$}, 
we can trace the detailed dynamics of this transition. 
One can observe that the field distribution already exhibits a mild local preference before the transition, 
implying that the wall configurations start forming even earlier. 
However, during this stage, the wall energy is negligible compared with the total energy, 
which is dominated by gradient and kinetic terms. 

{
Indeed, the emerging domain walls exhibit an approximately layered, chain-like structure 
along the direction perpendicular to the walls, if one neglects their curvature locally. 
The wall locations are characterized by 
\beq 
\phi_{\rm DW}/f_\phi ={ \{\cdots, -3\pi, -\pi, \pi, 3\pi, \cdots\}}
\subset [-\delta\phi_{\rm amp}/f_\phi, \, \delta\phi_{\rm amp}/f_\phi].
\eeq 
Note that domain walls can only annihilate with those possessing the same $\phi_{\rm DW}/f_\phi$. 
For illustration, consider an axion wave with wavelength $2\pi/k_{\rm typ}$ in the $z$-direction 
and amplitude $\delta\phi_{\rm amp}$, 
\beq
\phi(x)\propto \delta\phi_{\rm amp} \cos(k_{\rm typ} z),
\eeq
where we neglect the time-dependence. 
In this case, the field repeatedly crosses adjacent vacua as a function of $z$, and a chain of nearly parallel domain walls is formed. 
The number of walls within a half wavelength $\pi /k_{\rm typ}$ is approximately 
$\sim 2 \delta\phi_{\rm amp} / (2\pi f_\phi)$. 
As long as the gradient and kinetic energies dominate over the potential energy, the wall tension is subdominant compared to the wave energy density, so the walls do not enter a scaling regime but instead approximately follow the underlying axion-wave pattern. 
}

{
When such a chain of walls closes into a loop, it forms a layered, Baumkuchen-like domain-wall configuration. 
We refer to these layered loops as ``Baumkuchen domain walls''. 
The energy density of a Baumkuchen domain-wall loop can be estimated 
by assuming a single wall within a box of size 
$(\pi/k_{\rm typ}) \times 2\pi f_\phi / (2 \delta\phi_{\rm amp})$. 
We then obtain
\beq
\rho_{\rm BDW}
  \sim \frac{8 f_\phi^2 m_\phi}
  {2\pi^2 f_\phi / (2 k_{\rm typ} \delta\phi_{\rm amp})}
  = \frac{8}{\pi^2} f_\phi m_\phi \, \delta\phi_{\rm amp} \, k_{\rm typ}.
\eeq
This energy density remains smaller than the gradient and kinetic energies 
as long as $m_\phi^2 f_\phi^2 \lesssim k_{\rm typ}^2 \delta\phi_{\rm amp}^2$. 
}

At the transition ($a \sim 2.5$ in Fig.~\ref{fig:2}), 
the energy stored in the Baumkuchen domain walls becomes significant, 
and the wall tension drives the field distribution toward $\phi = 0$, 
the center of the initial condition. 
The Baumkuchen domain-wall network then approaches a scaling regime, 
containing $\mathcal{O}(1)$ large Baumkuchen domain wall per horizon, 
while smaller domain wall loops collapse. 
This evolution can be understood as a layered collapse of walls with different $\phi_{\rm DW}$ values. 
Such behavior can be seen in the snapshots of Fig.~\ref{fig:5} for $a = 3$ and $a = 5$, {where we also show the wall positions by the black contours}.
The longest surviving walls are {associated with} the domains containing the most populated vacuum, 
while smaller internal walls collapse earlier. 
The largest walls cannot collapse before the smaller ones due to 
the interaction forces between walls with different $\phi_{\rm DW}$. 

Then the domain walls with no superhorizon correlation 
(see Appendix~\ref{app1}) collapse shortly thereafter, 
completing the transition. 
After this collapse, axion particles are produced from the decay of the domain walls 
and subsequently behave as ordinary cold matter. 
At the moment of production, these particles are semi-relativistic, 
with typical momenta of order $m_\phi$, 
originating from the dynamics of the domain-wall collapse. 
This completes the overall picture of particle production and energy transfer
associated with this nonlinear transition. 

{
We also observe in the bottom panel of Fig.~\ref{fig:2} that, before the formation of the domain walls, the field
distribution becomes broader and exhibits a characteristic temporal
oscillation.  This originates from our ``nonrelativistic particle--like''
initial condition, in which the initial kinetic energy is slightly larger
than the gradient energy.  When the initial kinetic energy dominates, the positive- and
negative-frequency components of each mode acquire almost the same initial amplitude, 
and their interference produces a standing-wave component.  This leads to an
oscillatory exchange between the kinetic and gradient energies, as clearly
visible in the four right panels of Fig.~\ref{fig:1}.  The amplitude of this
oscillation is gradually reduced by cosmological redshift.  In this sense, the initially large kinetic energy drives the system into a regime where the kinetic and gradient contributions become balanced on
time average. When we impose an alternative, physically motivated initial condition that
corresponds to a massless ``propagating wave'' (i.e.\ $\omega = k$), the
left- and right-moving components no longer have correlated phases.  As a result, the standing-wave contribution is strongly suppressed and the oscillatory interference pattern disappears.  A detailed discussion and analysis as well as numerical simulation is
provided in Appendix~\ref{app:3}, and the phenomena found in the main part hold as well.
What we have shown is that, even if we impose an initial condition corresponding to nonrelativistic axion particles, the subsequent evolution behaves like a standing-wave radiation fluid and inevitably leads to the formation of Baumkuchen domain walls. 
}

One cannot obtain a matter-dominated epoch 
if the kinetic and gradient energies remain larger than the potential energy 
at redshifts $z < 5.5\times10^6$, 
even when the typical momentum is smaller than the mass. 
This once again confirms the condition given in \Eq{limit}.

\section{Conclusions and discussion}

We have found a generic limit for the minimal ALP dark matter~\Eq{limit} by considering generic field configurations. 
This is known if the dark matter is composed of the homogeneous mode. In this paper, we showed even if it is inhomogeneous, especially even when the typical momentum of the ALP is smaller than the mass, the limit holds as well. This is because if one violates the limit and requires the gradient and kinetic energies of the ALP much larger than the potential energy, it behaves as radiation irrelevant to the typical momentum of the ALP particles.

\begin{figure}[!t] 
    \begin{center}
        \includegraphics[width=105mm]{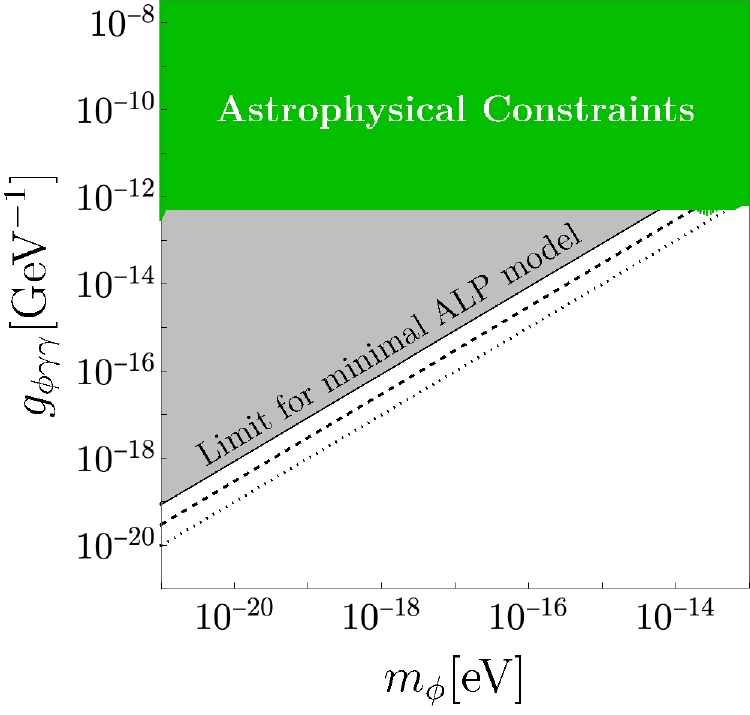}
    \end{center} 
    \caption{The generic limits derived from the minimal ALP dark matter model. The solid, dashed, and dotted lines correspond to the case for $c_\gamma =3, \, 1, \, 1/3$, respectively, from top to bottom. The green region indicates astrophysical constraints which are taken from the data compiled in Ref~\cite{AxionLimits}.}
    \label{fig:generic} 
\end{figure}

The ALP  is usually defined such that it couples to a pair of photons with the coupling
\beq
{\cal L} =c_\g \frac{\a}{4 \pi} \frac{\phi}{f_\f} F_{\mu \nu} \tilde F^{\mu \nu} 
\equiv \frac{1}{4} g_{\phi\g\g} \phi F_{\mu \nu} \tilde F^{\mu \nu}.
\eeq
We consider in the minimal setup $|c_\g|\lesssim 1$. This is because if $c_\g \gg 1$ we need mechanisms to enhance the photon coupling, e.g., coupling the ALP to many charged chiral fermions or mixing between multiple ALPs e.g.~\cite{Ho:2018qur,Cyncynates:2023esj, Murai:2024nsp}.\footnote{An example of the other type of loophole, which is not minimal, is to have another dark matter who gives the abundance to the ALP dark matter at late time Universe without changing much the equation of state. 
In this case, a more conservative limit may be, $m_\phi^2 f_\f^2<\rho_{\rm DM}^{\rm galaxy center}$, with $\rho_{\rm DM}^{\rm galaxy center}$ being the local dark matter density which is known to be dense such as around some galaxy centers or the center of the cuspy dwarf spheroidal galaxies. 
}

In this case, we get a generic cosmological upper limit on the ALP-photon coupling for the minimal model, which is shown in Fig.\,\ref{fig:generic} (See also \cite{Marsh:2019bjr,Dror:2020zru, Nakagawa:2022wwm}). 
In the case the ALP is discovered in the region violating the limit, it implies that the ALP is associated with a more complicated structure, which desires further model building and understanding to the relation to the fundamental law of physics. \\

We also find that when the typical ALP momentum is very small and the fluctuation amplitude is large, 
domain walls can form. 
These domain walls can be marginally long-lived and eventually collapse due to population bias. 
Such a process can generate gravitational waves, axion miniclusters, and primordial black holes. 
A further study on mechanisms that can produce large low-momentum modes—%
for example, through stimulated emission~\cite{Moroi:2020has, Moroi:2020bkq, Yin:2023jjj, Sakurai:2024apm}—%
would be an interesting direction (see also Ref.~\cite{Miyazaki:2025tvq}). 

Our findings also indicate that previous studies on ALP and QCD axion particle production~%
\cite{Moroi:2020has, Moroi:2020bkq, Nakayama:2021avl, Choi:2023jxw, Yin:2023jjj, Sakurai:2024apm, Co:2017mop, Harigaya:2019qnl} 
require certain revisions. 
In particular, if a significant number of ALPs are produced in low-momentum modes, %
where the potential energy is usually neglected,%
the matter--radiation transition of the ALP fluid is substantially modified. 
This has an important impact on the estimation of the dark matter abundance, 
which typically assumes that the comoving particle number or adiabatic invariant is conserved. 
However, our study shows that this assumption breaks down due to non-linear dynamics 
when $\rho_\phi$ redshifts to $m_\phi^2 f_\phi^2$, 
even if the typical momentum at that time is much smaller than $m_\phi$. 

It should be noted that the phenomena discussed so far are based on the assumption 
that the initial fluctuations do not possess superhorizon correlations~%
\cite{Gonzalez:2022mcx, Kitajima:2023kzu}, 
since we explicitly removed the IR modes in our numerical simulations. 
In this case, the ALP field $\phi$ does not serve as dark matter, 
because the superhorizon fluctuation—%
which is not an adiabatic mode—%
corresponds to an isocurvature perturbation. 
Nevertheless, such a setup can be naturally realized, for example, 
if the ALP is produced via the stimulated decay of a modulus field that remains light during inflation. 
As shown in Appendix~\ref{app1}, in this case the non-linear transition 
is accompanied by a long-lived network of domain walls. 
Therefore, the transition picture from radiation to matter discussed in this paper 
no longer applies; rather, the system undergoes a transition 
from radiation domination to a domain-wall--dominated phase. 
If the domain wall tension is too large, this would lead to a cosmological disaster. 
On the other hand, if the tension is sufficiently small, 
such scaling domain walls could naturally account for the hint 
of isotropic cosmic birefringence~%
\cite{Minami:2020odp, Takahashi:2020tqv, Kitajima:2022jzz, Diego-Palazuelos:2022dsq}. 
\\

{\it Note added:}~ While completing our study, we noticed a paper posted on arXiv~\cite{Harigaya:2025pox}, in which the authors discuss a similar bound to those in \cite{Marsh:2019bjr,Dror:2020zru,Nakagawa:2022wwm} and in the present work. 
The authors did not show that the axions with non-relativistic momenta can behave as radiation and did not point out the domain wall formation.

\section*{Acknowledgement}
This work is supported by JSPS KAKENHI Grant Nos.  20H05851 (W.Y.),  22K14029 (W.Y.), 22H01215 (W.Y.), Graduate Program on Physics for the Universe (Y.N.), and JST SPRING, Grant Number JPMJSP2114 (Y.N.). W.Y. is also supported by Incentive Research Fund for Young Researchers, and Selective Research Fund from Tokyo Metropolitan University.

\appendix 
\section{More on Baumkuchen domain wall}
\label{app2}
In the main text, we focused on a single simulation to study domain-wall dynamics. 
The setup there corresponds to an initial typical wavelength of $\sim 2\pi / K_{{\rm UV},i} \approx 5.6 / m_\star$ 
(cf.\ $1 / H_i =2 / m_\star$). 
Although this wavelength becomes shorter than the horizon size at the transition, 
one may wonder what happens if the initial wavelength is even shorter. 

\begin{figure}[!t]
    \begin{center}
        \includegraphics[width=145mm]{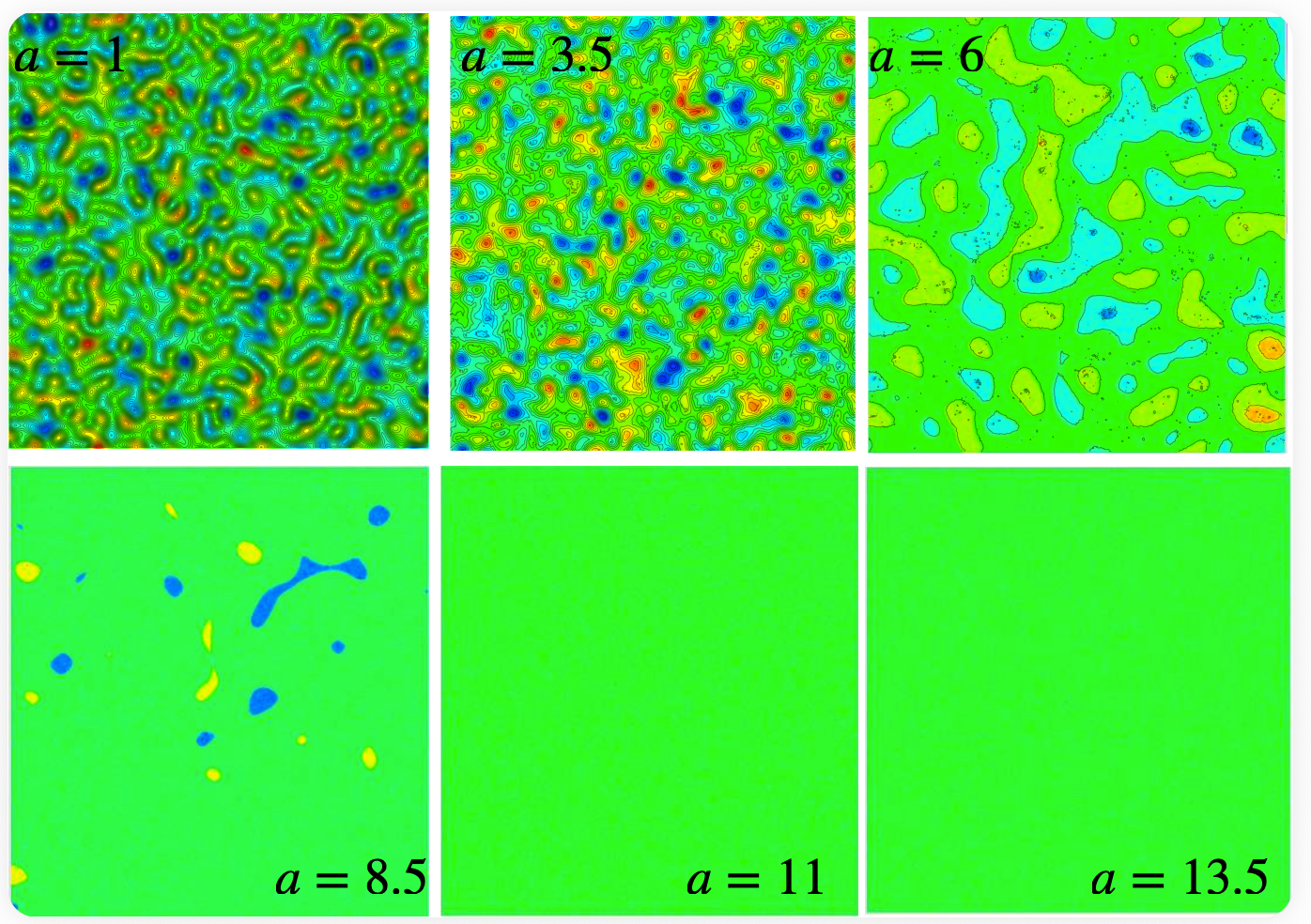}
        \vspace{-5mm}
    \end{center}
    \caption{
    Same as Fig.~\ref{fig:5}, but for the bottom-right panel of Fig.~\ref{fig:1}. }
    \label{fig:7}
\end{figure}

\begin{figure}[!t]
    \begin{center}
        \includegraphics[width=145mm]{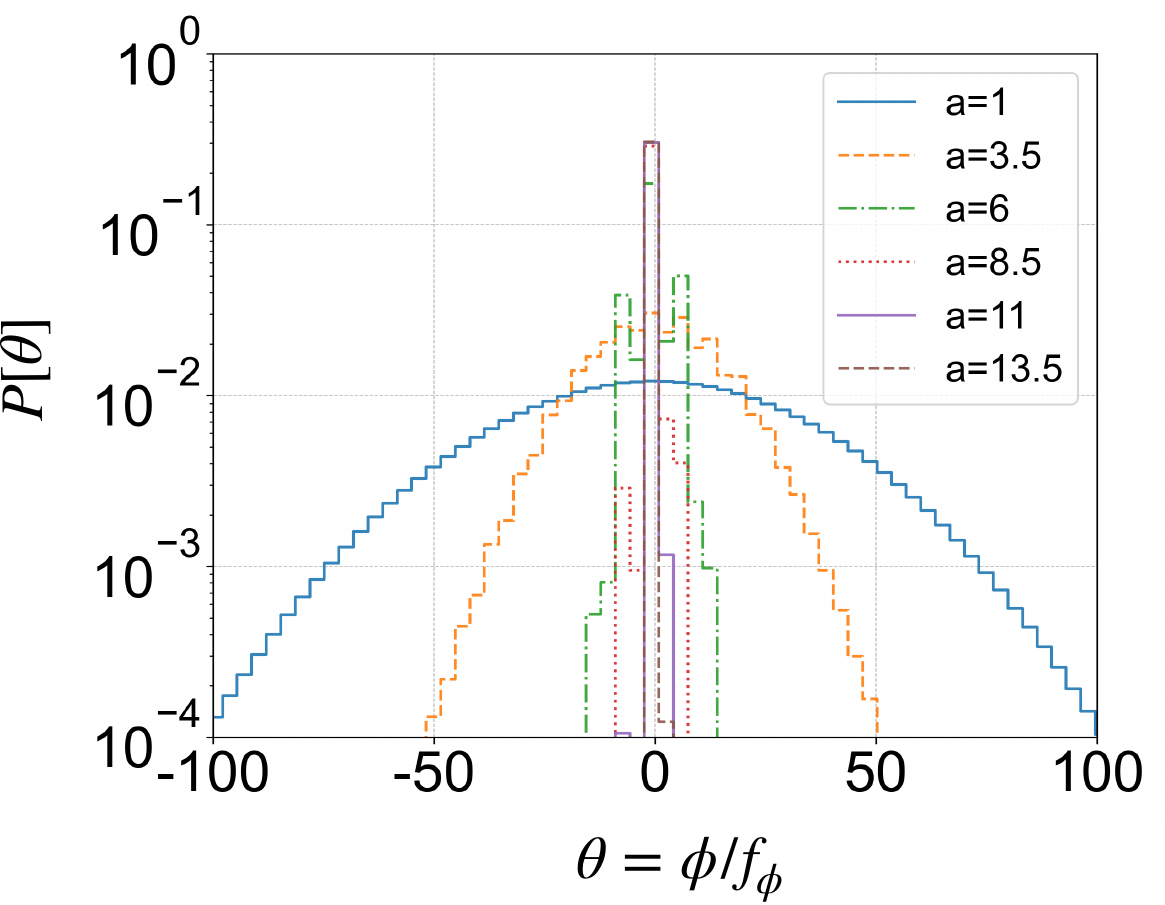}
        \vspace{-5mm}
    \end{center}
    \caption{
    Same as Fig.~\ref{fig:2}, but for the bottom-right panel of Fig.~\ref{fig:1} with $a=1,3.5,6,8.5,11,13.5$. }
    \label{fig:6}
\end{figure}

In this appendix, we analyze the data corresponding to the bottom-right panel of Fig.~\ref{fig:1}, 
which has a wavelength of $\sim 2.4 / m_\star$, comparable to the Hubble horizon from the beginning. 
The corresponding results shown in Figs.~\ref{fig:7} and~\ref{fig:6} are consistent with those described in the main text. 
In particular, the subhorizon {layer} of walls is almost frozen before the transition, 
as seen in the snapshots at $a = 1$ and $a = 3.5$ in Fig.~\ref{fig:7}. 
The wall layers nearly follow the wave due to the suppressed tension, as discussed earlier. 
Soon after the transition at $a \sim 4.4$, 
when the tension energy becomes significant, 
the small-inertia walls collapse and disappear, 
leaving only the large-scale walls in the scaling regime. 

At $a = 6$, the comoving horizon size is $a / H = 12\, m_\star^{-1}$, 
which is about $1/6$ of the box size ($\simeq 63 / m_\star$), 
consistent with the scaling wall length. 
The wall network subsequently disappears due to population bias and does not persist for long. 
At the final snapshot ($a = 11$), the domain walls have almost vanished, 
although the slice still contains about $3^2$ Hubble volumes. 
This behavior can also be seen in the distribution plot in Fig.~\ref{fig:6}. 

\section{{Long-lived domain walls from scale-invariant low-momentum modes}}
\label{app1}

Let us consider another setup by taking
\beq
K_{\rm IR} = 0, \qquad q = 0,
\eeq
which implies that the reduced power spectrum ${\cal P}_{\delta\phi}$ is scale independent, 
i.e., it corresponds to a scale-invariant fluctuation that typically arises during inflation. 
This setup essentially follows Refs.~\cite{Gonzalez:2022mcx, Kitajima:2023kzu}. 
The difference from those works is that here we adopt a cosine-type potential rather than a $\phi^4$ model. 
In the $\phi^4$ case, large-amplitude fluctuations tend to relax into white-noise--like configurations, 
whereas in the cosine potential the long-range correlation remains, as discussed in Ref.~\cite{Kitajima:2023kzu}. 

\begin{figure}[!t]
    \begin{center}
        \includegraphics[width=145mm]{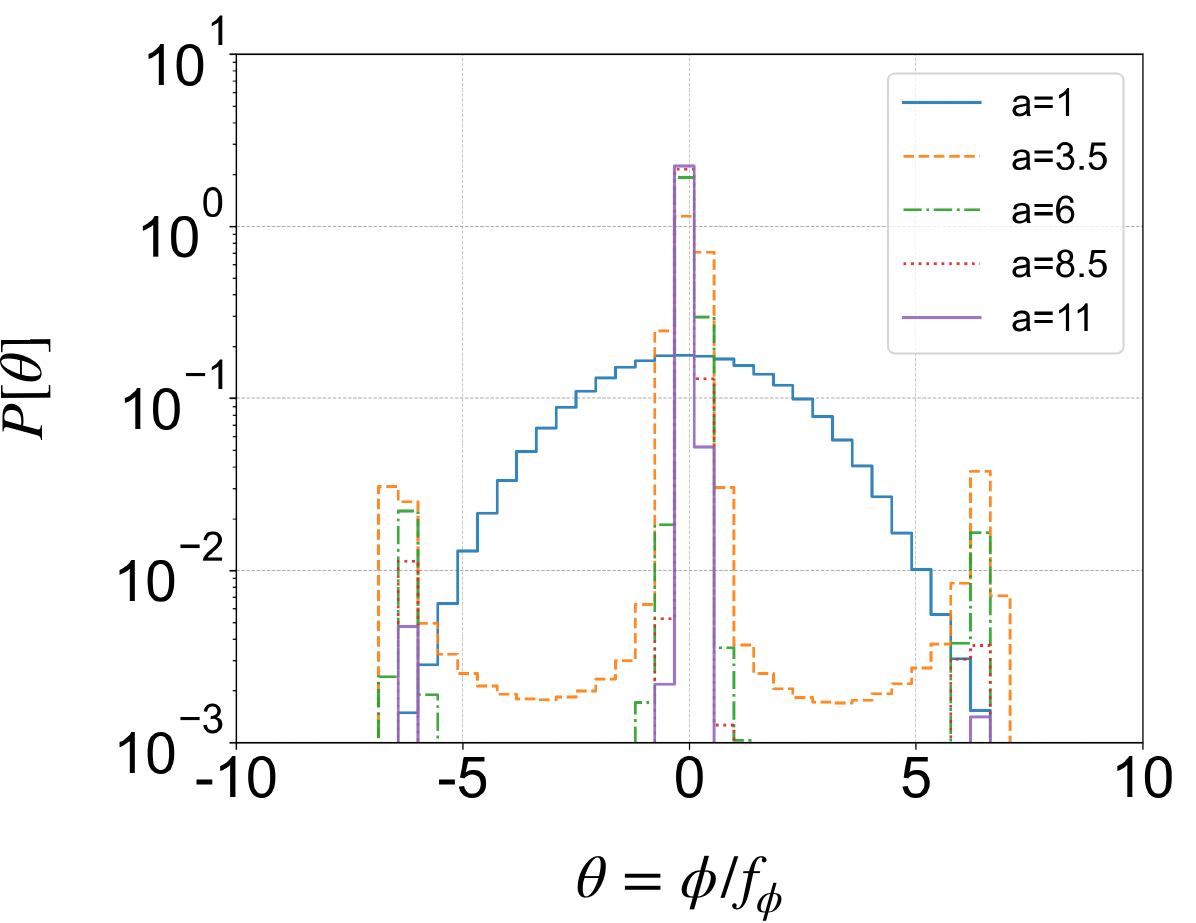}
        \vspace{-5mm}
    \end{center}
    \caption{
    Distribution of $\phi/f_\phi$ for the case with scale-invariant fluctuations.}
    \label{fig:3}
\end{figure}

The result is shown in Fig.~\ref{fig:3}, 
where we choose $K_{\rm UV}^2 = 0.3\,m_\phi^2$, 
$L \simeq 126/m_\star$, 
$m_\phi = m_\star$, 
and $\tilde{\phi} = 1.2 f_\phi$ 
with $1024^3$ lattices. 
These parameters are chosen such that the fluctuations are marginally at the transition threshold. 
Even in this case, the domain walls remain long-lived, 
which drastically changes the behavior compared to what we discussed in the main text. 

A more general situation can be realized with arbitrary $q$ 
but with both $K_{\rm UV}$ and $K_{\rm IR}$ smaller than the Hubble scale at the transition. 
In this case, large-scale domain walls form, 
and the ALP energy density undergoes a prolonged domain-wall epoch.

\section{{Initial conditions and propagating wave of massless ALP}}
\label{app:3}

An initial condition with \Eq{inw} with very small momentum implies that the kinetic energy
dominates over the gradient energy,
\begin{equation}
(\dot{\phi}_k)^2 \sim m_\phi^2 \phi_k^2 \;>\; k^2 \phi_k^2,
\end{equation}
which is often motivated in the context of nonrelativistic particle-like excitations.
We adopt it in the main text because we wish to demonstrate that even nonrelativistic
axion particles behave effectively as radiation whenever their energy
density is larger than the potential height.  With such an initial condition,
the kinetic and gradient components of the energy naturally oscillate with
each other.

This behavior follows from the equation of motion with the potential term
neglected:
\begin{equation}
\ddot{\phi}_{\vec k} + k^2 \phi_{\vec k} + 3 H \dot{\phi}_{\vec k} \sim 0.
\end{equation}  
For $H \ll k\equiv |\vec k|$, the solution is
\begin{equation}
\phi_{\vec{k}} \sim 
c_+\, e^{i\!\int^t dt'\, k}\, a^{-1}
\;+\;
c_-\, e^{-i\!\int^t dt'\, k}\, a^{-1},
\end{equation}
with the redshifting physical momentum, $k$.

If the initial condition at $t=0$ and $a=1$ satisfies 
$\lvert \dot{\phi}_{i,\vec k} \rvert \gg \lvert k\, \phi_{i,\vec k} \rvert$, 
one finds 
\begin{equation}
|c_+ - c_-| \gg |c_+ + c_-|.
\end{equation}
For instance, the dispersion relation adopted in the main text gives
$c_+ - c_- \sim (m_\phi/k)(c_+ + c_-)$, which leads to $c_+ \simeq c_-$.
The mode therefore contains a large standing-wave component.
This standing-wave interference drives an oscillatory exchange between the
kinetic and gradient energies, producing the broadening and oscillation of
the field distribution seen in the bottom panel of Fig.~\ref{fig:2}
for $a=1$--$3$ when the initial gradient energy is subdominant and
$\rho_\phi \gg m_\phi^2 f_\phi^2$.

\medskip

We can also consider an alternative, physically motivated initial condition
by replacing
\begin{equation}
\omega = \sqrt{k^2 + m_\phi^2}
\qquad\longrightarrow\qquad
\omega = k .
\end{equation}
This choice is justified by the fact, shown in this paper, that an axion behaves as radiation
whenever $\rho_\phi \gg m_\phi^2 f_\phi^2$, regardless of the instantaneous
value of $k/m_\phi$.  Such conditions naturally arise when relativistic ALPs
are produced in the early Universe, with their momenta subsequently
redshifting below $m_\phi$.  
If one aims to simulate the evolution near the transition
$\rho_\phi \sim m_\phi^2 f_\phi^2$, adopting $\omega=k$ is appropriate, since
the kinetic and gradient energies start comparable and both redshift as
radiation (and because following the entire cosmological history in a
lattice simulation is numerically impractical).

Since $c_\pm$ are randomly generated in the simulation, it is typical that,
for each momentum mode, one of $c_\pm$ dominates.  The standing-wave
interference is therefore suppressed, and the energy density
$\propto |\phi_k|^2$ shows little temporal oscillation.  
The resulting configuration is effectively a propagating axion wave.

\begin{figure}[!t]
    \begin{center}
        \includegraphics[width=145mm]{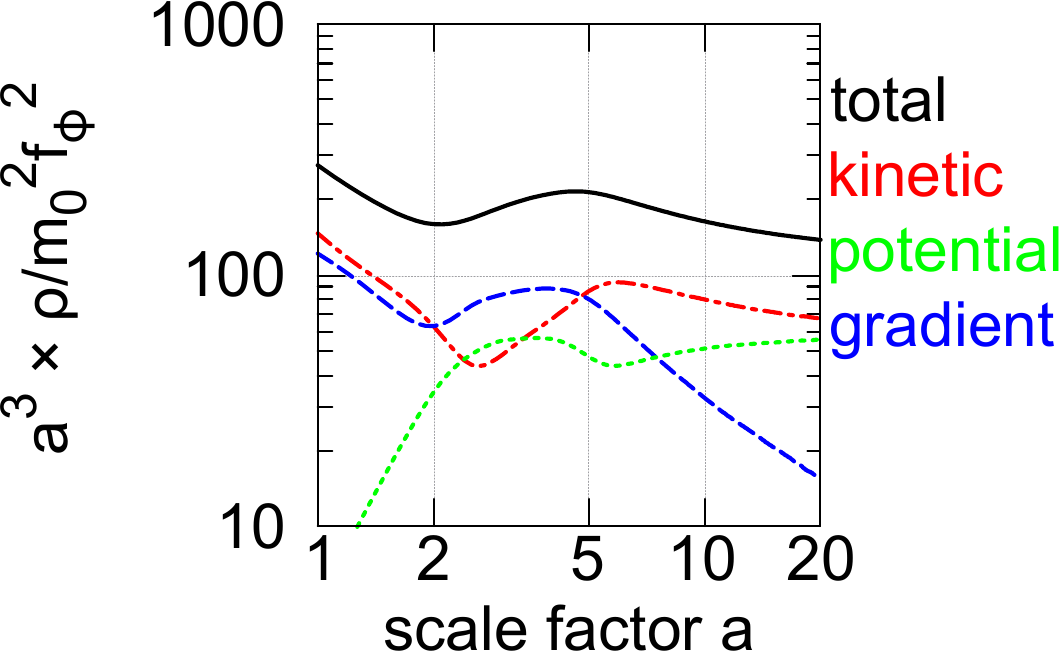}
        \vspace{-5mm}
    \end{center}
    \caption{
    Same as the bottom-left panel of Fig.\,\ref{fig:1}, but adopting the initial condition $\omega = k$.}
    \label{fig:10}
\end{figure}

\begin{figure}[!t]
    \begin{center}
        \includegraphics[width=145mm]{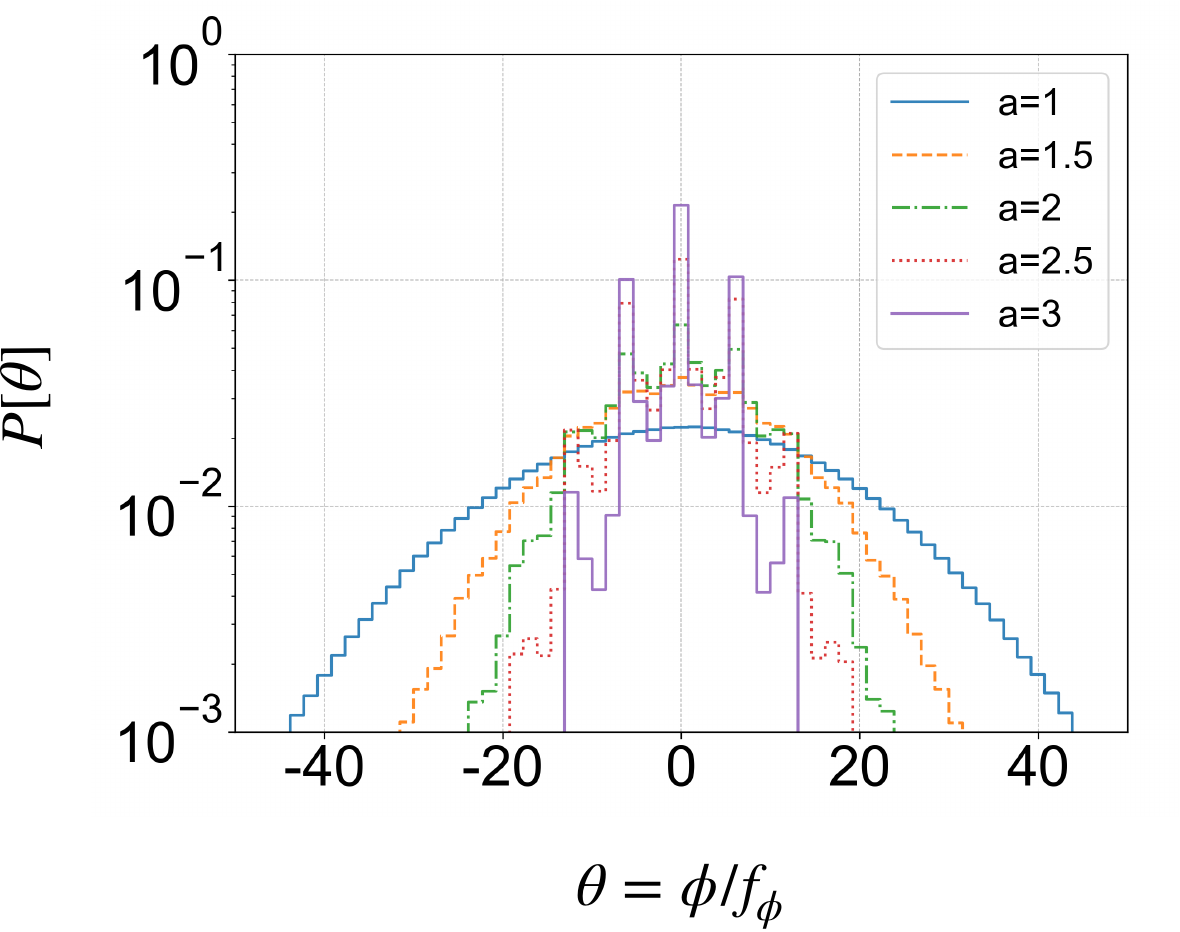}
        \vspace{-5mm}
    \end{center}
    \caption{
    Distribution of $\phi/f_\phi$ corresponding to Fig.\,\ref{fig:10}.}
    \label{fig:11}
\end{figure}

The numerical results for this initial condition are shown in
Figs.~\ref{fig:10} and \ref{fig:11}, corresponding to the bottom-middle
panel of Fig.~\ref{fig:1} and the bottom panel of Fig.~\ref{fig:2},
respectively.  As expected, the oscillatory behavior is strongly suppressed,
while the transition timescale---analytically $a\simeq 2.5$ estimated from the initial total energy---remains consitent with the result.

Although this initial condition is not suitable for explicitly demonstrating
the statement that ``nonrelativistic axions behave as radiation,'' since
the initial state is relativistic, it is more consistent with cosmological
production and avoids unphysical standing-wave oscillations.  For these
reasons we regard $\omega=k$ as the appropriate initial condition for
future simulations.

\bibliography{GenericALPDMbound.bib}
\end{document}